\documentclass[12pt]{iopart}
\usepackage[dvips]{graphicx}
\usepackage{latexsym}
\usepackage{iopams}
\usepackage{verbatim,times,bbm}

\begin{document}

\title{Qubit-portraits of qudit states and quantum correlations}

\author{C. Lupo$^{1}$, V. I. Man'ko$^{2}$, G. Marmo$^{1}$}

\address{$^1$ Dipartimento di Fisica dell'Universit\`a di Napoli `Federico II'
and Istituto Nazionale di Fisica Nucleare (INFN) Sezione di Napoli,
Complesso Universitario di Monte Sant'Angelo, via Cintia, Napoli,
I-80126, Italy}
\address{$^2$ P. N. Lebedev Physical Institute, Leninskii Prospect 53, Moscow 119991, Russia}

\eads{\mailto{lupo@na.infn.it}, \mailto{manko@sci.lebedev.ru},
\mailto{marmo@na.infn.it}}

\begin{abstract}

The machinery of qubit-portraits of qudit states, recently
presented, is consider here in more details in order to characterize
the presence of quantum correlations in bipartite qudit states. In
the tomographic representation of quantum mechanics, Bell-like
inequalities are interpreted as peculiar properties of a family of
classical joint probability distributions which describe the quantum
state of two qudits. By means of the qubit-portraits machinery a
semigroup of stochastic matrices can be associated to a given
quantum state. The violation of the CHSH inequalities is discussed
in this framework with some examples, we found that quantum
correlations in qutrit isotropic states can be detected by the
suggested method while it cannot in the case of qutrit Werner
states.

\end{abstract}

\pacs{03.65.-w, 03.67.-a}

\section{\label{intro}Introduction}

Entanglement is probably one of the most intriguing and fascinating
characteristic of quantum mechanics \cite{Schro}, its importance
lies at the heart of the physical interpretation of the theory. The
scientific interest and efforts towards the understanding and a
complete characterization of entanglement is motivated both by its
role in the conceptual foundation of quantum theory and by all the
recent proposals and applications which lead to consider
entanglement as a resource for quantum information and computation
tasks \cite{N-C}.

Although the two concepts are not equivalent, the presence of
entanglement is strongly related to quantum non-locality. The
fundamental tools to study quantum non-locality, i.e. quantum
correlations, are the Bell-like inequalities. A violation of a
Bell-like inequality is an evidence of the presence of non-local
correlations in the quantum state. It is well known that only
entangled states can violate Bell-like inequalities. In the present
paper we study bipartite mixed states entanglement by looking at
violations of a Bell-like inequality, to do this we exploit the
point of view given by the tomographic description of quantum
mechanics \cite{tomog}.

The main goal of the present contribution is to further analyze the
linear map which defines the \emph{qubit-portraits} of qudit state
introduced in \cite{Chernega}. In particular we consider how this
map can be used to describe quantum correlation in a bipartite
quantum system. This paper has two main ingredients: the first is
the tomographic description of quantum mechanics, the second is
related to the CHSH inequalities \cite{Bell,CHSH}. The tomographic
approach is known to be mathematically equivalent to the other
descriptions of quantum mechanics based, for instance, on density
matrices or Wigner functions. Nevertheless there are two
conceptually relevant differences: the first one is that in the
tomographic approach one deals only with well defined (classical)
probability distributions which are directly related to
experimentally accessible relative frequencies of measurement
outcomes; the second one is that, in order to define a tomogram, one
needs additional information about the observables related to a
given experimental setup. From these considerations it can be argued
that the tomographic approach can be viewed as a rather natural
framework to study Bell-like inequalities. In the present paper we
study the well known CHSH inequalities in this framework. Although
the tomographic description of quantum mechanics can be defined in
full generality \cite{Ventriglia}, here we concentrate our attention
on quantum systems with finite levels.

Among a plethora of proposed criteria to detect entanglement, a
prominent position is held by a family of methods which are based on
the action of special linear maps on the set of separable quantum
states. Examples are the criteria based on positive but not
completely positive maps \cite{Horo} (like the criterion of the
positive partial transpose \cite{Peres}) and the realignment
criterion \cite{RC} which can be understood from a unique point of
view based on linear contractions \cite{L_C}. Another example is
given by the criterion based on partial scaling transform
\cite{partial} which is a linear map that is neither completely
positive nor positive. In the present paper we make use of the
qubit-portraits of a qudit state \cite{Chernega} which is again a
linear map but is defined in the tomographic description of quantum
mechanics.

The paper is organized in the following way. In section
\ref{tomo-intro} we briefly recall some definitions and basic
properties about tomograms. In section \ref{Bell-tomo} the CHSH
inequalities are presented in the framework of the tomographic
approach to quantum mechanics. In section \ref{portraits} the
machinery of qubit-portraits of a qudit system is considered in
order to deal with higher dimensional systems, examples for qubit
and qutrit Werner and isotropic states are presented. The paper ends
with final remarks and conclusions in section \ref{theend}.

\section{\label{tomo-intro}Introduction to quantum tomograms}

Let us consider a $d$-level quantum system with the associated
Hilbert space $\mathcal{H}\cong C^d$ and a chosen basis
$\{|m\rangle\}_{m=1,\dots,d}$. Given a state of a system expressed
by means of a density operator $\rho$, there are several ways to
define a corresponding tomogram; let us first consider the
definition of \emph{unitary tomogram}. The diagonal elements
$\langle m | \rho | m \rangle$ of the density operator are the
populations in the given basis, they constitute a well defined
probability distribution. The knowledge of the populations in a
given basis is in general not sufficient to reconstruct the off
diagonal elements of the density operator, on the other hand the
knowledge of the populations in all possible bases gives complete
information about the quantum state of the system. As the unitary
group acts transitively on the family of bases, a generic basis
$\{|m'\rangle\}$ can be identified with a special unitary
transformation $u\in \mathrm{SU}(d)$ with $|m'\rangle=u|m\rangle$.
These considerations yield to the definition of the unitary tomogram
as follows:
\begin{equation}\label{unitary}
\omega_\rho(m,u)\equiv\langle m|u^\dag\rho u|m\rangle.
\end{equation}
The tomogram is thus a family of well defined probability
distributions over $d$ possible measurement outcomes, which depends
on the $d^2-1$ parameters defining a special-unitary transformation.
It is thus apparent that the tomogram explicitly gives the
probability distributions for the outcomes of all the possible
projective measurements allowed by the principles of quantum
mechanics. As a matter of fact this is a redundant description, a
lower number of bases would be sufficient as long as they constitute
a tomographic set \cite{Ventriglia}.

Let us now consider a special case, in which the $d$-level system is
indeed a spin-$j$ particle, with $d=2j+1$, and the state vectors
belonging to the basis are eigenstates of the angular momentum along
a quantization axis, say $\hat{z}$. In this case, one can be mostly
interested in measurements of polarization along a generic direction
$\hat{n}$. Hence one is led to define the \emph{spin tomogram} ad
follows:
\begin{equation}\label{spin}
\omega^j_\rho(m,D)\equiv\langle m|D^\dag\rho D|m\rangle,
\end{equation}
where $D$ belongs to a spin-$j$ irreducible representation of the
group $\mathrm{SU}(2)$ and has the following expression (see
\cite{Landau}, for instance):
\begin{equation}
\langle m' | D | m \rangle = e^{-im'\phi} d^j_{m'm}(\theta)
e^{-im\gamma},
\end{equation}
where
\begin{equation}
\fl d^j_{m'm}(\theta) = \left[ \frac{(j+m)!(j-m)!}{(j+m')!(j-m')!}
\right]^{1/2} \left( \sin{\frac{\theta}{2}} \right)^{m-m'} \left(
\cos{\frac{\theta}{2}} \right)^{m+m'}
P_{j-m}^{(m-m',m+m')}(\cos{\theta})
\end{equation}
is the Wigner matrix and $P_{j-m}^{(m-m',m+m')}$ are the Jacobi
polynomials. An unitary operator $D$ is uniquely identified by the
three Euler angles, nevertheless since only the diagonal elements of
the (rotated) density operator appear in the definition, the
tomogram depends only on two Euler angles, say $\theta$ and $\phi$,
or equivalently on a point on the Bloch sphere $\hat{n}\equiv
(\sin{\theta}\cos{\phi},\sin{\theta}\sin{\phi},\cos{\theta})$.
Notice that both kinds of tomograms are mathematically equivalent to
the density matrix description of the quantum states. In the case of
spin tomography an additional physical information is added, this
information allows to restrict to bases generated by an irreducible
representation of $\mathrm{SU}(2)$ acting on a properly chosen
fiducial one.

Let us study quantum entanglement in the tomographic picture (see
also some aspects of this approach in \cite{Andr-}). In order to set
properly the problem of separability of a quantum state, one needs
primarily to identify a partition of the whole system into a number
of subsystems each of dimension $d_k$. This can be done
mathematically with the only constraint that $\Pi d_k=d$,
nevertheless the definition of subsystems is in general physically
determined and depends on the experimentally achievable observables
and operations. To fix the ideas, let us for instance consider the
case of a spin-$j$ particle which turns to be a bipartite system
composed of a spin-$j_1$ and spin-$j_2$, with $d_k=2j_k+1$ and $d_1
d_2=d$. It is natural to define another kind of tomogram, which we
call \emph{local spin tomogram}, as follows:
\begin{equation}\label{local-spin}
\omega^{j_1 j_2}_\rho(m_1,m_2,D_1,D_2)\equiv\langle m_1 m_2
|D_1^\dag\otimes D_2^\dag\rho D_1\otimes D_2|m_1 m_2\rangle,
\end{equation}
where $m_k=-j_k,-j_k+1,\dots,j_k$, and $D_k$ are unitary irreducible
representations of $\mathrm{SU}(2)$. An analogous construction can
be made for \emph{local unitary tomography}, which yields to the
definition:
\begin{equation}\label{local-unitary}
\omega_\rho(m_1,m_2,u_1,u_2)\equiv\langle m_1 m_2 |u_1^\dag\otimes
u_2^\dag\rho u_1\otimes u_2|m_1 m_2\rangle.
\end{equation}
These definitions may be immediately extended to the multi-partite
case. Notice that, while in the density matrix description the
information about the internal structure of the system has to be
inserted as an additional information, in the tomographic approach
it is included in the chosen kind of tomogram from the very
beginning. In the case of local spin tomography (\ref{local-spin})
the tomogram is a family of probability distributions depending on
the two pairs of Euler angles $(\theta_k,\phi_k)$ which determine
the directions of polarization $\hat{n}_1$ and $\hat{n}_2$ for the
first and second particle respectively.

Since the tomogram is a family of well defined probability
distributions we find that the tomographic approach to quantum
mechanics can be a natural candidate to deal with quantum
probabilities and correlations and also to study violation of
Bell-like inequalities. Let us consider an observable $X$, it
identifies a preferred basis $|\bar{m}\rangle=\bar{u}|m\rangle$ in
terms of its eigenstates, then the expectation value is simply
written as
\begin{equation}
\langle X \rangle_\rho = \sum_m x_m \omega_\rho(m,\bar{u}),
\end{equation}
where $x_m$ are the corresponding eigenvalues. Let us consider the
case of a bipartite system with a couple of local observables $X_1$
and $X_2$ with corresponding eigenstates
$|\bar{m}_k\rangle=\bar{u}_k|m_k\rangle$ and eigenvalues $x_{m,k}$.
In the tomographic picture the correlation $C_\rho(X_1,X_2)=\langle
X_1 X_2\rangle_\rho$ is written as follows:
\begin{equation}
C_\rho(X_1,X_2)=\sum_{m_1,m_2}
x_{m_1,1}x_{m_2,2}\omega_\rho(m_1,m_2,\bar{u}_1,\bar{u}_2).
\end{equation}
Given a bipartite system with simply separable density operator
$\rho=\rho_1\otimes\rho_2$ it follows from the definitions
(\ref{local-spin}) or (\ref{local-unitary}) that
\begin{equation}
\omega_\rho(m_1,m_2,u_1,u_2)=\omega_{\rho_1}(m_1,u_1)\omega_{\rho_2}(m_2,u_2),
\end{equation}
that is, the tomogram itself is the product of two tomograms and, in
particular, it defines a family of uncorrelated joint probability
distributions. By linearity, it follows that a generic separable
state with density matrix $\rho=\sum_k p_k \rho_1^k\otimes\rho_2^k$
has a tomogram of the form
\begin{equation}\label{separable}
\omega_\rho(m_1,m_2,u_1,u_2)=\sum_k p_k
\omega_{\rho_1^k}(m_1,u_1)\omega_{\rho_2^k}(m_2,u_2),
\end{equation}
which corresponds to a family of probability distributions with
(classical) correlations. Notice that the tomogram is a family of
well defined \emph{classical} probability distributions in
\emph{any} case, for separable states the decomposition
(\ref{separable}) exists with constant $p_k \geq 0$ and
$\omega_{\rho_1^k}(m_1,u_1)$ and $\omega_{\rho_2^k}(m_2,u_2)$ which
are well defined tomograms.

\section{\label{Bell-tomo}CHSH inequalities in the tomographic
picture}

In this section we review the CHSH inequalities exploiting the
tomographic description of quantum mechanics and quantum
correlations. In order to do this, we introduce a stochastic matrix
which is determined by a given tomogram whose structure is related
to the form of the CHSH inequalities. These inequalities were
introduced in \cite{CHSH} as a generalization of the original Bell's
inequalities \cite{Bell} in order to relax some experimentally
unfeasible assumptions. The setting in which the inequalities are
formulated is made by an ensemble of pairs of correlated particles
moving in opposite directions and entering respectively two
measurement apparatus, say $I_a$ and $II_b$, where $a$ and $b$ are
adjustable parameters defining the apparatus configuration. At each
side of the experiment a dichotomic observable is measured, say
$A(a)$ for the apparatus $I_a$ and $B(b)$ for the apparatus $II_b$.
The choice of the observables depends on the value of the
\emph{local} parameters $a$ and $b$, each of the \emph{local}
observable is taken to have as possible outcomes $+1$ and $-1$.  The
correlation function between the two observables is
$C_\rho(a,b)=\langle A(a)B(b) \rangle_\rho$, in the hypothesis of
local realism the following inequalities hold
\begin{equation}\label{Bell-CHSH}
B = |C_\rho(a,b)+C_\rho(a,c)+C_\rho(d,b)-C_\rho(d,c)|\leq 2
\end{equation}
for any value of the parameters $a, b, c, d$ and any $\rho$.

In order to describe these inequalities from the point of view of
the tomographic representation, we define an associated matrix in
terms of which the inequalities (\ref{Bell-CHSH}) can be written,
eventually this matrix will turn to be a stochastic matrix. Let us
first consider the simplest case of a bipartite system composed by
two two-level systems. In order to deal with the generic case, we
consider the unitary tomogram corresponding to the density matrix
$\rho$:
\begin{equation}
\omega_\rho(m_1,m_2,a,b)
\end{equation}
where $a$ and $b$ are short hand notations for $u_1(a)$ and
$u_2(b)$. Putting $m=1$ and $m=-1$ respectively for polarization
parallel and anti-parallel to the quantization direction we can
define the following matrix:
\begin{eqnarray}\fl
M_\rho=\left[\begin{array}{cccc}
\omega(1,1,a,b) & \omega(1,1,a,c) & \omega(1,1,d,b) & \omega(1,1,d,c) \\
\omega(1,-1,a,b) & \omega(1,-1,a,c) & \omega(1,-1,d,b) & \omega(1,-1,d,c) \\
\omega(-1,1,a,b) & \omega(-1,1,a,c) & \omega(-1,1,d,b) & \omega(-1,1,d,c) \\
\omega(-1,-1,a,b) & \omega(-1,-1,a,c) & \omega(-1,-1,d,b) &
\omega(-1,-1,d,c)
\end{array}\right]
\end{eqnarray}
Notice that each column of this matrix is a well defined probability
distribution which corresponds to the tomogram with particular
values of the parameters, hence $M$ is a \emph{stochastic matrix}.
Thus a stochastic matrix is associated to a quantum tomogram in a
way which is somehow analogous to the relation between density
matrices and quantum maps \cite{Jam,Alicki,Jam_infty}. Also notice
that the order in which the columns are organized with respect to
the parameters $a,b,c,d$ resembles the structure of a direct
product. It is easy to check that for simply separable states the
associated stochastic matrix factorizes as the direct product of two
stochastic matrices each one corresponding to one-particle tomogram:
\begin{eqnarray}\fl
\rho=\rho_1\otimes\rho_2 \ \ \Rightarrow \ \
M=\left[\begin{array}{cc}
\omega_1(1,a) & \omega_1(1,d) \\
\omega_1(-1,a) & \omega_1(-1,d)
\end{array}\right]\otimes
\left[\begin{array}{cc}
\omega_2(1,b) & \omega_2(1,c) \\
\omega_2(-1,b) & \omega_2(-1,c)
\end{array}\right].
\end{eqnarray}
That is, a simply separable state corresponds to a factorized
stochastic matrix. Analogously, a separable state corresponds to a
stochastic matrix which is the convex sum of factorized stochastic
matrices.

With the labeling $m=-1,1$ the discrete index in the tomogram is
just the value of the relevant observable, so the expectation value
for the correlation is simply written as $C(u_1,u_2)=\sum_{m_1,m_2}
m_1 m_2 \omega(m_1,m_2,u_1,u_2)$. Introducing the matrix
\begin{eqnarray}
I=\left[\begin{array}{cccc}
1 & -1 & -1 & 1 \\
1 & -1 & -1 & 1 \\
1 & -1 & -1 & 1 \\
-1 & 1 & 1 & -1
\end{array}\right]
\end{eqnarray}
the CHSH inequalities (\ref{Bell-CHSH}) can be written in the
following way:
\begin{equation}\label{I-ineq}
B = |\mathrm{tr}(IM)| \leq 2.
\end{equation}
This expression will be used in the following sections where we
define, by means of the machinery of the \emph{qubit-portraits}
introduced in \cite{Chernega}, a stochastic matrix in the case of a
bi-partite system composed of two qudits.

\section{\label{portraits}Qubit-portraits of qudit systems}

In this section we consider the CHSH inequalities in the case of a
system composed of two qudits. In order to do this one needs to
define a couple of dichotomic observables and to study the
correlations between them. This discussion belongs to a general
setting made of a system composed of $2$ ($d$-dimensional)
subsystems; in each of one, $2$ local observables are measured and
each measurement has $2$ possible outcomes. While in the qubit case
any non trivial observable can be associated with a dichotomic
observable with outcomes $+1$ and $-1$, this is not the case for
qudit systems in which dichotomic observables do not represent the
generic case. This kind of problem was already considered in
\cite{Lupo}, in the present work we exploit the machinery introduced
in \cite{Chernega} which allows one to define a family of
probability distributions which mimic a qubit tomogram and give a
complete description of a qudit system, this kind of representation
is called \emph{qubit-portrait}.

As we have already recalled, the tomogram of a quantum state is a
family of probability distributions over all possible measurement
outcomes in a given basis, where each measurement outcome
corresponds to a one-dimensional projector $P(m)=|m \rangle\langle
m|$. In the same way one can consider a two-dimensional or in
general a $n$-dimensional projector defined as
$P(m_0,m_1,\dots,m_{n-1})=\sum_{k=0}^{n-1} |m_k\rangle\langle m_k|$
and consider the corresponding probability. Since the projectors on
the basis vectors are orthogonal to each other, in the tomographic
representation this probability is given by the sum over independent
events $\sum_k \omega(m_k,u)$. As an examples, let us consider the
case of a qutrit system. In this case we have an unitary tomogram
$\omega(m,u)$ where $m=0,1,2$ and $u\in\mathrm{SU}(3)$. Identifying
the \emph{events} $m=0$ and $m=1$, we can define a
\emph{qubit-portrait} of the qutrit state as the family of
probability distribution $\omega'(m',u)$, with $m'=0,1$ and
$\omega'(0,u)=\omega(0,u)+\omega(1,u)$ and
$\omega'(1,u)=\omega(2,u)$. Analogously, one can define other two
qubit-portraits of the qutrit state. In the same way we can reduce
any qudit tomogram to a family of probability distribution over a
dichotomic variable and so define a qubit-portraits representation
for any qudit tomogram. The same considerations can be extended to
the case of spin tomography and to the case of multipartite systems:
for instance, a tomogram for the state of a system composed of two
qutrits can be reduced to a family of probability distributions over
two dichotomic variables, which corresponds to a two-qubit portrait
of the two-qutrit system. In this fashion one can define, as in the
previous section, a (square) stochastic matrix using the qubits
portraits of a qudit-qudit system.

\section{\label{qutrits}Qubit-portraits of qutrit states and CHSH inequalities}

In this section we study the CHSH inequalities applied to the case
of qutrit-qutrit system, in particular we focalize our attention
onto the families of Werner states \cite{Werner} and isotropic
states \cite{isot}. In order to define a dichotomic variable, we
reduce the qudit states to a family of probability distributions
which are the corresponding qubit-portraits. This yields to identify
a stochastic matrix, analogous to the one presented in section
\ref{Bell-tomo}, which is defined by means of the qubit-portrait.
Having reduced the qudit-qudit system to an \emph{effective}
qubit-qubit system, we can consider the inequalities (\ref{I-ineq}).
In principle one can write several inequalities (not all
independent) which correspond to all the possible qubit-portraits
that can be defined starting from the given qudits tomogram. In the
case of qubit-qubit system, the CHSH inequalities have been already
considered in the tomographic picture in \cite{Lupo}, now we
consider the inequalities (\ref{I-ineq}) defined with the help of
qubit-portraits machinery.

The first family is given by the qudit-qudit Werner states, which is
a one parameter family of quantum states defined as:
\begin{equation}\label{Werner-state}
W=(d^3-d)^{-1}\left[ (d-\phi)\mathbbm{I}+(d\phi-1)\mathbbm{V}
\right]
\end{equation}
for $\phi\in[-1,1]$, where $\mathbbm{I}$ is the identity operator in
the two-qudit space and $\mathbbm{V}$ is the \emph{flip operator}
defined as
$\mathbbm{V}|\psi_1\rangle|\psi_2\rangle=|\psi_2\rangle|\psi_1\rangle$.
The state (\ref{Werner-state}) is separable for $\phi \geq 0$ and
entangled otherwise. The second family is given by the isotropic
states:
\begin{equation}\label{isotropic-state}
S = (d^2-1)^{-1}\left[ (1-p)\mathbbm{I} +
(pd^2-1)|\psi\rangle\langle\psi| \right]
\end{equation}
for $p\in[0,1]$, where
$|\psi\rangle=\frac{1}{\sqrt{d}}\sum_{i=1}^d|ii\rangle$ is a
maximally entangled states. The state (\ref{isotropic-state}) is
separable for $p \leq d^{-1}$ and entangled otherwise. Notice that,
for $d=2$ and $p \geq 0$, the two families are related by a
re-parametrization and a partial transposition.

We also need to specify what kind of tomogram we want to use, for
the sake of simplicity we restrict our discussion to the case of
polarization measurements, this is the case in which we take the
subgroup of the unitary group $\mathrm{U}(d)$ given by an
irreducible representation of $\mathrm{SU}(2)$ and consider the
local spin tomogram. In this case the parameters defining the local
observables are just the pair of Euler angles which identify the
direction of polarization. In the two-qubit case, the local group is
$\mathrm{SU}(2)\otimes\mathrm{SU}(2)$ acting on
$\mathbbm{C}^2\otimes\mathbbm{C}^2$, when the qubit portraits arise
from qutrit one should use $\mathrm{SU}(2)\subset\mathrm{SU}(3)$
acting irreducibly.

\begin{figure}
\centering
\includegraphics[width=0.5\textwidth]{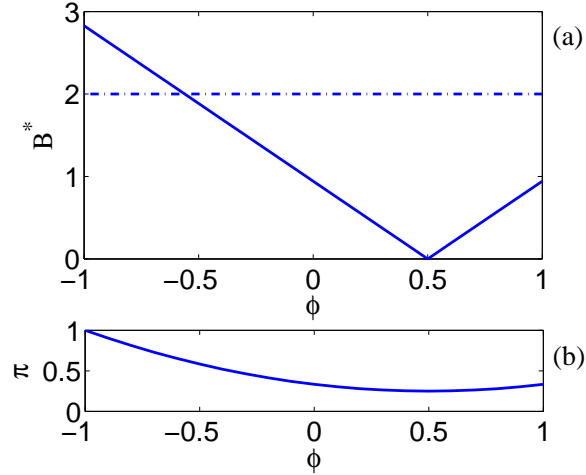}
\caption{(a) Maximum value of the Bell number (\ref{I-ineq}) for a
two qubit Werner state (\ref{Werner-state}) as a function of the
state parameter (solid line) compared with the maximum value allowed
by local hidden variables theories (dot-dashed line). (b) Purity of
the two qubit Werner state.} \label{werner_qubit}
\end{figure}

\begin{figure}
\centering
\includegraphics[width=0.5\textwidth]{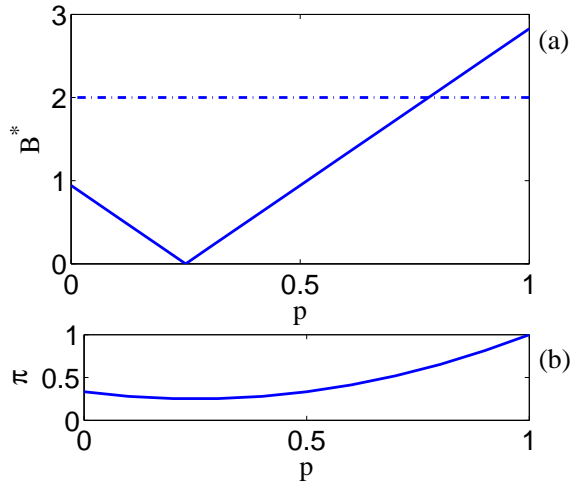}
\caption{(a) Maximum value of the Bell number (\ref{I-ineq}) for a
two qubit isotropic state (\ref{isotropic-state}) as a function of
the state parameter (solid line) compared with the maximum value
allowed by local hidden variables theories (dot-dashed line). (b)
Purity of the two qubit isotropic state.} \label{iso_qubit}
\end{figure}

First of all, let us consider the case of qubits. In this case the
construction of the qubit portraits is redundant and our discussion
is just a different way to deal with CHSH inequalities, nevertheless
this example can be an useful term of comparison with respect to
higher dimensional non trivial configurations. We have computed the
spin tomogram of the two qubit Werner states and computed the
maximum of the quantity (\ref{I-ineq}), denoted $B^*$. Notice that
the maximum is taken with respect to all the possible choices of
local observables which in the case of spin tomography are
identified by four unit vectors on the Bloch sphere $\hat{n}_a,
\hat{n}_b, \hat{n}_c, \hat{n}_d$. The results are shown in figure
\ref{werner_qubit}a.
The same calculation has been done for the case of two-qubit
isotropic states and the corresponding results are plotted in figure
\ref{iso_qubit}b. The plots \ref{werner_qubit}b and \ref{iso_qubit}b
show the purity $\pi = \tr\rho^2$ as a function of the parameter of
the corresponding states.

For the case of two qutrit state, we have compared the results
obtained with the qubit-portrait method with the qutrit-qutrit
Bell's inequalities presented in \cite{collins} which generalize the
CHSH inequalities. In our notation we can write them as:
\begin{eqnarray}
I_3 & = & \left\{ P[A(a)=B(b)] + P[A(c)=B(b)-1]  + \right. \nonumber\\
& & \left. P[A(c)=B(d)] + P[A(a)=B(d)]\right\} + \nonumber\\
& & - \left\{ P[A(a)=B(b)-1] + P[A(c)=B(b)] \right. + \nonumber\\
& &  \left. P[A(c)=B(d)-1] + P[A(a)=B(d)+1]\right\} \leq 2,
\label{compare}
\end{eqnarray}
where
\begin{equation}
P([A(a) = B(b) + k] \equiv \sum_j \omega(j + k,j,a,b),
\end{equation}
and the sum $j+k$ is modulo $3$. Notice that, also in this case, the
inequalities can be written using the language of tomograms in a
natural way.

\begin{figure}
\centering
\includegraphics[width=0.5\textwidth]{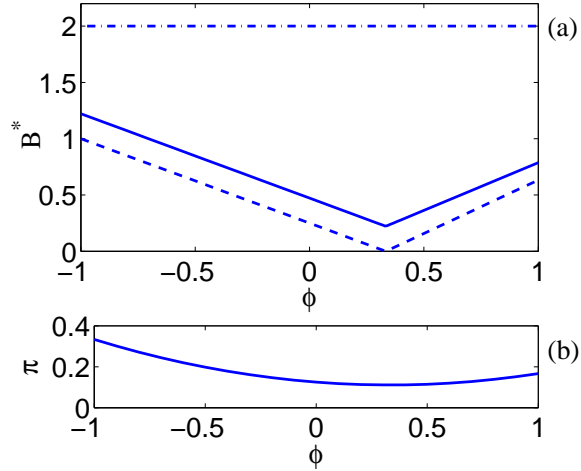}
\caption{(a) Maximum value of the Bell number (\ref{I-ineq}) for a
two qutrit Werner state (\ref{Werner-state}) as a function of the
state parameter (solid line) compared with the maximum value of
$I_3$ from equation (\ref{compare}) (dashed line) and the maximum
value allowed by local hidden variables theories (dot-dashed line).
(b) Purity of the two qutrit Werner state.} \label{werner_qutrit}
\end{figure}

\begin{figure}
\centering
\includegraphics[width=0.5\textwidth]{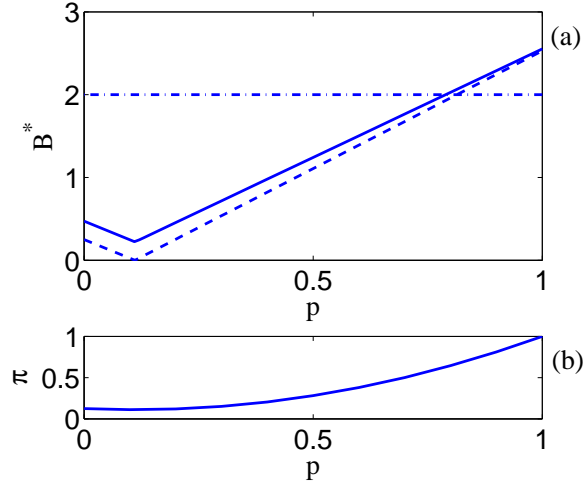}
\caption{(a) Maximum value of the Bell number (\ref{I-ineq}) for a
two qutrit isotropic state (\ref{isotropic-state}) as a function of
the state parameter (solid line) compared with the maximum value of
$I_3$ from equation (\ref{compare}) (dashed line) and the maximum
value allowed by local hidden variables theories (dot-dashed line).
(b) Purity of the two qutrit isotropic state.} \label{iso_qutrit}
\end{figure}

For the two-qutrit Werner states, we have first considered all the
possible two-qubit portraits which are computed by means of the
procedure described in section (\ref{portraits}). The maximum of the
Bell number (\ref{I-ineq}) is determined with respect to both
polarization vectors which define the set of local observables and
the different qubit-portraits of the two qutrit system. The results
are plotted in figure \ref{werner_qutrit}a together with the maximum
value of the analogous quantity $I_3$ from equation (\ref{compare}).
The plot shows that the qubit-portraits method cannot reveal quantum
correlations in two qutrit Werner states.
Analogously we have computed the maximum of (\ref{I-ineq}) for the
two-qutrit isotropic states, in this case, as shown in figure
\ref{iso_qutrit}a, the qubit-portrait method is able to witness the
presence of quantum correlations. The results are plotted together
with the maximum value of $I_3$ from equation (\ref{compare}).

The different capability of the CHSH inequalities approached with
the machinery of the qubit-portraits to recognize quantum
correlations in Werner and isotropic state for $d=2,3$ can be
related to the different value of the purity of the corresponding
states which are plotted in figure \ref{werner_qutrit}b and
\ref{iso_qutrit}b.

For the case of the two-qutrit isotropic states, the minimal value
of the state parameter, arising from our method, that yields to a
violation of the Bell's inequalities is $p_{\mathrm{min}}^B \simeq
0.7893$ for the CHSH inequalities and $p_{\mathrm{min}}^{I_3} \simeq
0.8139$ for the inequalities (\ref{compare}). In our notation, the
singlet fraction is $q = (9p-1)/8$, yielding to $q_{\mathrm{min}}^B
\simeq 0.7630$ and $q_{\mathrm{min}}^{I_3} \simeq 0.7906$. A
comparison with the results presented in \cite{collins,quNit} yields
to the observation that the local spin tomography, although based on
an irreducible action of the group $\mathrm{SU}(2)$, cannot give
complete information about violation of Bell inequalities. In other
words, even though a tomographic set gives complete information
about the quantum state and allows the reconstruction of the density
operator, the local spin tomogram cannot necessary reach the
configuration corresponding to the maximal violation of a Bell-like
inequality.

\section{\label{theend}Conclusions and outlook}

In this paper we have further investigated the method of
qubit-portraits of qudit states first discussed in \cite{Chernega}.
This method arises in a natural way in the tomographic description
of quantum mechanics, it allows to map a qudit tomogram onto a
family of probability distributions which mimic a family of qubit
tomograms. The method is applied in relation to the study of
non-classical correlations in quantum systems, it allows a study of
the CHSH inequalities for generic bipartite qudit systems.

Exploiting the tomographic approach to quantum mechanics, it is
possible to associate a stochastic matrix to any bipartite quantum
system with a finite number of levels, its structure is related to
the structure of the CHSH inequalities and in term of it the
presence of quantum correlations can be studied. Some examples have
been presented regarding two special classes of bipartite states,
namely Werner and isotropic states. The results show that performing
the operation of the qubit-portraits can lead to some loss of
information about quantum correlations , as it is witnessed by the
absence of violations of the CHSH inequalities in the case of the
qubit-portraits of two qutrits Werner states. On the other hand, the
study of other two-qutrit Bell's inequalities with the framework of
quantum tomography, leads to the conclusion that even though the
spin tomogram allows the reconstruction of the quantum state, it
does not necessary provide the maximal violation of a Bell-like
inequalities.

Following \cite{Ventriglia,Jam_infty} in future publications we will
consider possible extensions of the present work to the case of
systems with higher dimensions and continuous variables. Other
possible applications of the qubit-portraits method can be the study
of other Bell-like inequalities which involve more than two choices
of local observables per part.

\ack V.~I.~Man'ko thanks the University of Napoli `Federico II' and
I.N.F.N. sezione di Napoli for kind hospitality.


\section*{References}

\begin{thebibliography}{99}

\bibitem{Schro} Schr\"odinger E 1935
{\it Naturwissenschaften} {\bf 23} 807, 823, 844

\bibitem{N-C} Nielsen M A and Chuang I L 2000
{\it Quantum Computation and Quantum Information} (Cambridge:
Cambridge University Press)

\bibitem{tomog} Mancini S, Man'ko V I and Tombesi P 1996
{\it \PL A} {\bf 213} 1;
%
Dodonov V V and Man'ko V I 1997
{\it ibid.} {\bf 229} 335;
%
Man'ko V I and Man'ko O V 1997
J.~Exp.~Ther.~Phys.~{\bf 85} 430

\bibitem{Chernega} Chernega V N, Man'ko V I 2007
J.~Russ.~Laser Res. {\bf 28} 2

\bibitem{Bell} Bell J S 1965
{\it Phys.} {\bf 1} 195

\bibitem{CHSH} Clauser J F, Horne M A, Shimony A, Holt R A 1969
{\it \PRL} {\bf 23} 880

\bibitem{Ventriglia} Man'ko V I, Marmo G, Simoni A, Stern A,
Sudarshan E C G and Ventriglia F 2005
{\it \PL A} {\bf 351} 1;
%
Man'ko V I, Marmo G, Simoni A, Ventriglia F 2006
{\it Open~Sys.~Inf.~Dyn.} {\bf 13} 239

\bibitem{Horo} Horodecki M, Horodecki P and Horodecki R 1996
{\it \PL A} {\bf 223} 1

\bibitem{Peres} Peres A 1996
{\it \PRL} {\bf 77} 1413

\bibitem{RC} Chen K, Wu L -A 2003
{\it Quant.~Inf.~Comp.} {\bf 3} 193;
%
Rudolph O 2002 Further results on the cross norm criterion for
separability {\it Preprint} quant-ph/0202121

\bibitem{L_C} Horodecki M, Horodecki P and Horodecki R 2006
{\it Open~Sys.~Inf.~Dyn.} {\bf 13} 103

\bibitem{partial} Man'ko O V, Man'ko V I, Sudarshan E C G, Zaccaria F 2006
{\it \PL A} {\bf 357} 255;
%
Man'ko O V, Man'ko V I, Marmo G, Shaji A, Sudarshan E C G, Zaccaria
F 2005
{\it ibid.} {\bf 339} 194;
%
Lupo C, Man'ko V I, Marmo G, Sudarshan E C G 2005
{\it \JPA} {\bf 38} 10377

\bibitem{Andr-} Andreev V A, Man'ko V I, Man'ko O V, Shchukin
E V 2006
{\it Theor.~Math.~Phys.} {\bf 146} 172

\bibitem{Landau} Landau L D, Lifshitz E M 1977
{\it Quantum Mechanics: Non-Relativistic Theory} (Oxford: Pergamon
Press) 3rd edition

\bibitem{Jam} Sudarshan E C G, Mathews P M and Rau J 1961
{\it \PR} {\bf 121} 920;
%
Jamiolkowski A 1972
%
{\it Rep.~Math.~Phys.} {\bf 3} 275

\bibitem{Alicki} Alicki R, Lendi K 1987
{\it Quantum Dynamical Semigroups and applications} (Berlin:
Springer Verlag)

\bibitem{Jam_infty} Marmo G, Asorey M, Kossakowski A, Sudarshan E C G 2005
{\it Open~Sys.~Inf.~Dyn.} {\bf 12} 319

\bibitem{Lupo} Lupo C, Man'ko V I, Marmo G 2006
{\it \JPA} {\bf 39} 12515

\bibitem{Werner} Werner R F 1989
{\it \PR A} {\bf 40} 4277

\bibitem{isot} Horodecki M, Horodecki P 1999
{\it \PR A} {\bf 59} 4206;
%
Vollbrecht K G H, Werner R F 2001
{\it ibid.} {\bf 64} 062307

\bibitem{collins} Collins D, Gisin N, Linden N, Massar S, Popescu S
2002
{\it \PRL} {\bf 88} 040404

\bibitem{quNit} Chen J- L, Kaszlikowski D, Kwek L C, Zukowski M and
Oh C H 2002
{\it \PR A} {\bf 64} 052109;
%
Kaszlikowski D, Gnacinski P, Zukowski M, Miklaszewski W, Zeilinger A
2000
{\it \PRL} {\bf 85} 4418

\end{thebibliography}
\end{document}